\documentclass[showpacs,preprintnumbers,amsmath,amssymb,twocolumn]{revtex4}

\usepackage{graphicx}
\usepackage{dcolumn}
\usepackage{bm}
\newcommand{\be}{\begin{equation}}
\newcommand{\ee}{\end{equation}}
\begin{document}
\title{Mesoscopic lattice Boltzmann modeling of flowing soft systems}

\author{Roberto Benzi}\affiliation{University of Tor Vergata and INFN, via della Ricerca Scientifica 1, 00133 Rome, Italy}
\author{Sergio Chibbaro}\affiliation{Istituto Applicazioni Calcolo, CNR, V.le del Policlinico 137, 00161, Rome, Italy}
\author{Sauro Succi}\affiliation{Istituto Applicazioni Calcolo, CNR, V.le del Policlinico 137, 00161, Rome, Italy}

\begin{abstract}
A mesoscopic multi-component lattice Boltzmann model with short-range repulsion
between different species and short/mid-ranged attractive/repulsive interactions 
between like-molecules is introduced.
The interplay between these composite interactions gives rise to
a rich configurational dynamics of the density field, exhibiting
many features of disordered liquid dispersions (micro-emulsions) and soft-glassy materials, such as
long-time relaxation due to caging effects, anomalous enhanced viscosity,  
ageing effects under moderate shear and flow above a critical shear rate.
\end{abstract}

\maketitle


The rheology of flowing soft systems, such as emulsions, foams, gels, slurries, colloidal glasses
and related fluids, is a fast-growing sector of modern non-equilibrium thermodynamics, 
with many applications in material science,chemistry 
and biology~\cite{Rus_89}.
These materials exhibit a number of distinctive features, such
as long-time relaxation, anomalous viscosity, aging behaviour, whose quantitative
description is likely to require profound extensions of non-equilibrium statistical mechanics. 
The study of these phenomena sets a pressing challenge for computer simulation as well,
since characteristic time-lenghts of disordered fluids can escalade tens of decades
over the molecular time scales.
To date, the most credited techniques for computational studies of these complex flowing materials
are Molecular Dynamics and Monte Carlo simulations~\cite{Allen}.  
Molecular dynamics in principle provides a fully ab-initio description of the system, but 
it is limited to space-time scales significantly shorter than experimental ones. 
Monte Carlo methods are less affected by these limitations, but they
are bound to deal with equilibrium states. 
As a result, neither MD nor MC can easily take into account the non-equilibrium
dynamics of complex flowing materials, such as micro-emulsions, on space-time scales of hydrodynamic interest.  
In the last decade, a new class of mesoscopic methods, based on minimal
lattice formulations of Boltzmann's kinetic equation, have captured significant
interest as an efficient alternative to continuum methods
based on the discretization of the Navier-Stokes equations for non-ideal
fluids \cite{Ben_92}. 
To date, a very popular such mesoscopic technique is the so-called 
pseudo-potential-Lattice-Boltzmann (LB) method, developed over a decade 
ago by Shan and Chen (SC) \cite{SC_93}.
In the SC method, potential energy interactions are represented
through a density-dependent mean-field pseudo-potential, $\Psi[\rho]$, and 
phase separation is achieved by imposing a short-range attraction between the
light and dense phases. 
In this Letter, we provide the first numerical evidence that a suitably
extended, two-species, mesoscopic lattice Boltzmann model
is capable of  reproducing many features of soft-glassy (micro-emulsions), such as 
structural arrest, anomalous viscosity, cage-effects and ageing under shear. 
The key feature of our model is the capability to investigate the rheology  of these systems 
on space-time scales of hydrodynamic interest.

The kinetic lattice Boltzmann equation takes the following form \cite{Ben_92}: 
\begin{equation}
f_{is}(\vec{r} + \vec{c}_i, t + \Delta t) -  f_{is}(\vec{r} , t ) =  -\frac{\Delta t}{\tau_s}[f_{is}(\vec{r} , t )-f_{is}^{(eq)}(\vec{r} , t )]   + F_{is}\Delta t
\label{eq:be}
\end{equation}
where $f_{is}$ is the probability of finding a particle of species $s$ 
at site $\vec{r}$ and time $t$, moving along the $i$th 
lattice direction defined by the discrete speeds $\vec{c}_i$ with $i=0,...,b$. 
The left hand-side of (\ref{eq:be}) stands for molecular free-streaming, whereas the right-hand side 
represents the time relaxation (due to collisions) towards local Maxwellian equilibrium on a time
scale $\tau_s$ and $F_{is}$ represents the volumetric body force due to 
intermolecular (pseudo)-potential interactions. 
\begin{figure}
\vspace{0.5cm}
\includegraphics[scale=0.45]{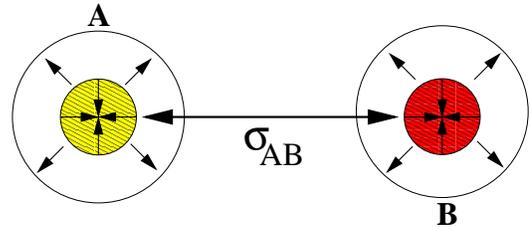}
\caption{The two components A and B interact via a repulsive pseudo-potential, which supports a surface tension $\sigma_{AB}$. 
Moreover, each component experiences an attractive interaction in the first Brillouin zone and a repulsive one 
acting on both Brillouin zones. Each of these interactions can be tuned through a separate coupling constant.}
\label{Fig:1}
\end{figure}
\begin{figure}
\vspace{0.5cm}
\hspace{0cm}
\includegraphics[height=4cm,width=4cm]{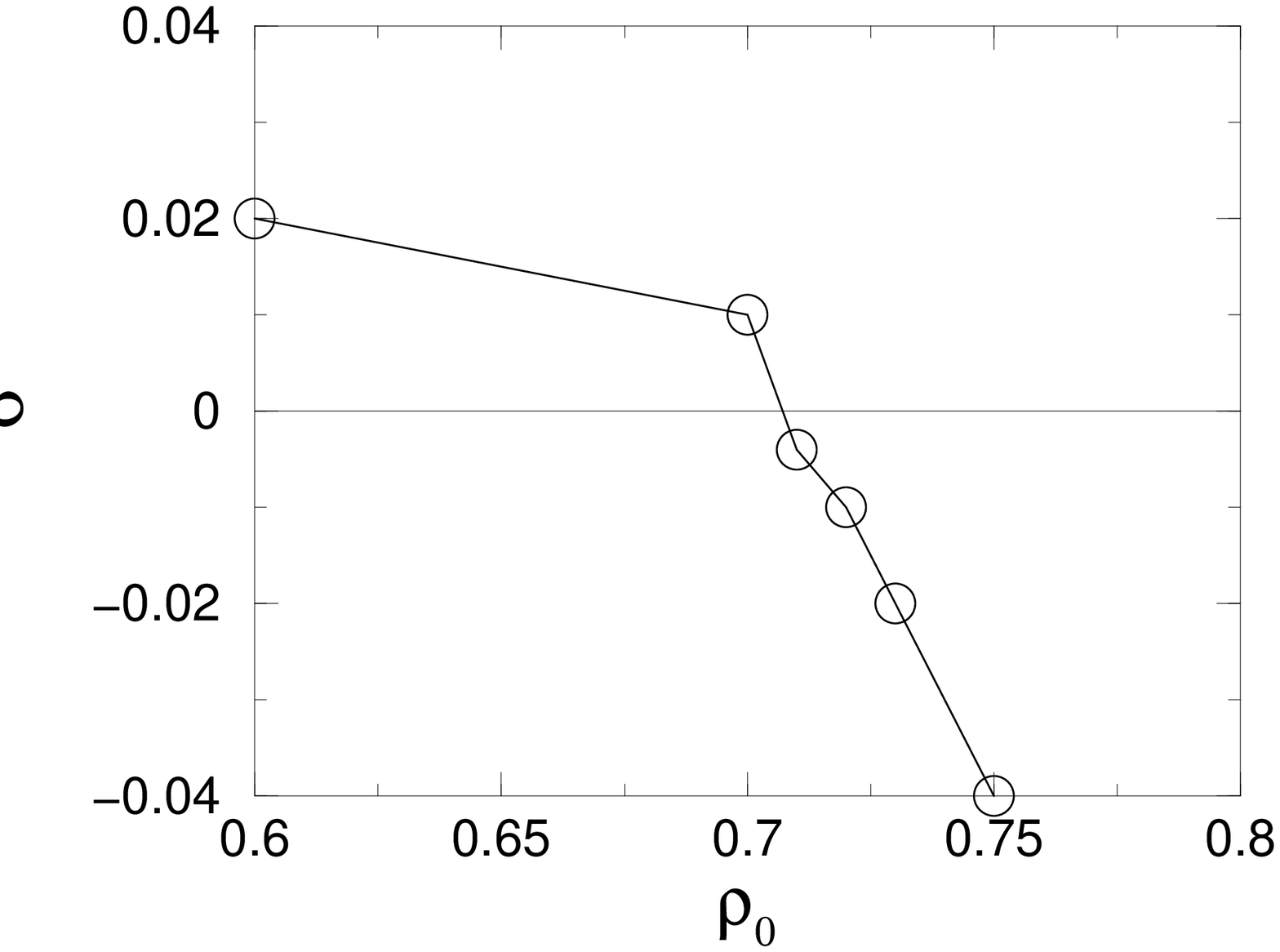}\vspace{+0.cm}
\includegraphics[height=4cm,width=4cm]{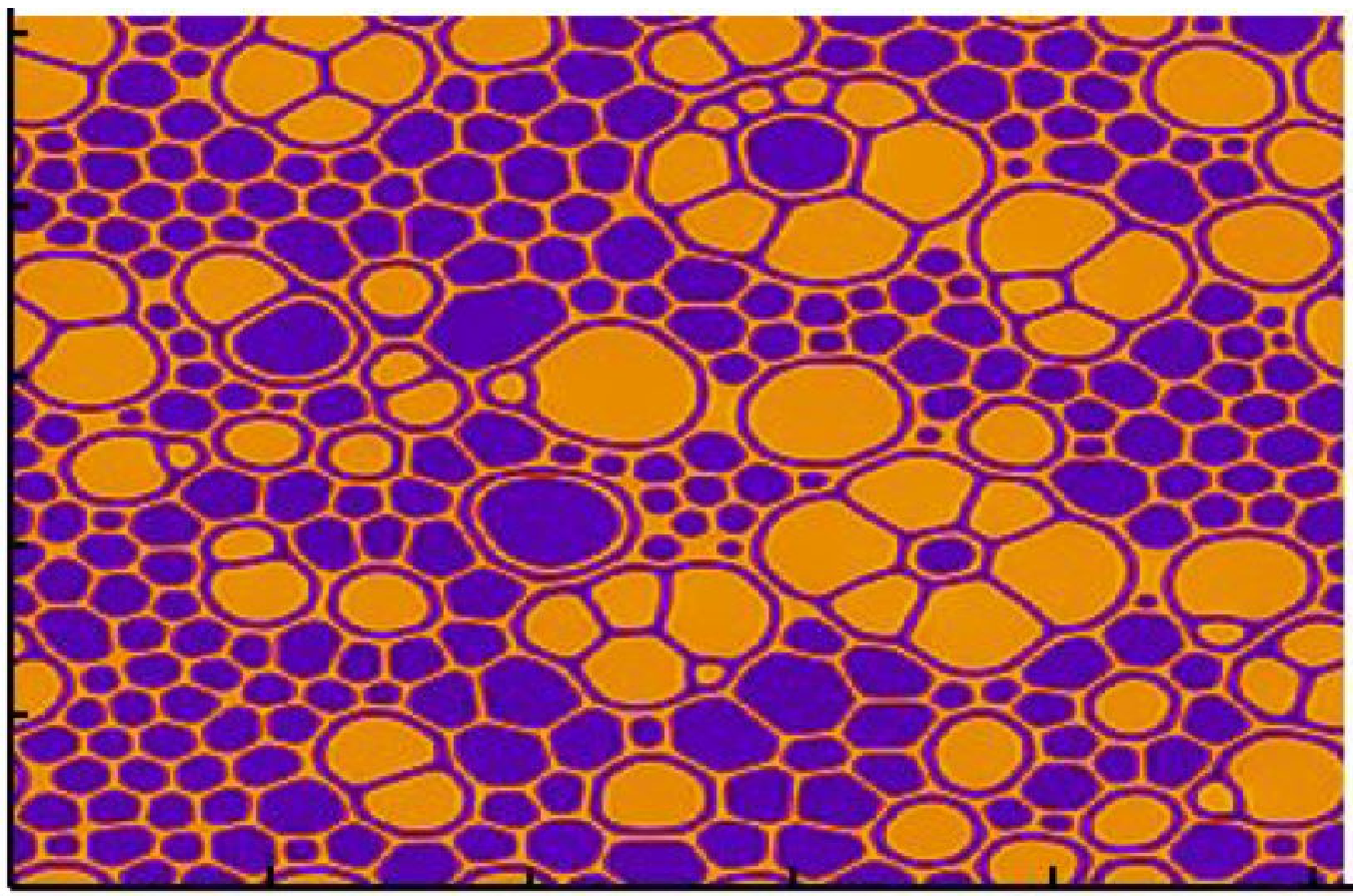}\vspace{+0cm}
\caption{
The nominal surface tension of the two-component fluid as a function of the reference density $\rho_0$ (left panel).
The coupling parameters are $G_{A1}=-12.55$, $G_{A2}=11.70$, $G_{B1}=-11.80$, $G_{B2}=10.95$, $G^{X}=0.58$. 
By increasing the reference density $\rho_0$, the relative strength
of interspecies repulsion is weakened as compared to other interactins, thus leading to a 
net decrease of the surface tension. 
Above a critical value, $\rho_{crit} \sim 0.72$, the nominal surface tension turns negative.
Right panel: A typical snapshot of the density field $\rho_A$ at time $2\times210^6$ (lattice units) at resolution $512^2$.
Because of the small but positive value of $\sigma$, the system proves capable of supporting 
fairly complex metastable density configurations.}
\label{Fig:2}
\end{figure}
The pseudo-potential force within each species consists of an attractive component,
acting only on the first Brillouin region (belt), and a repulsive one acting on both belts, whereas 
the force between species is short-ranged and repulsive: 
$\vec{F}(\vec{r}, t) = \vec{F}^a_s(\vec{r}, t) + \vec{F}^r_s(\vec{r}, t)+\vec{F}^X_s$, where
\begin{eqnarray}
\label{FORCE}
\vec{F}^a_s(\vec{r}, t) &=& G^a_s \Psi_s(\vec{r};t) \sum_{i=0}^{b_1} w_i \Psi_s(\vec{r}_{1i},t) \vec{c}_{1i} \Delta t,  \nonumber\\ 
\vec{F}^r_s(\vec{r}, t) &=& 
  G^r_{s} \Psi_s(\vec{r};t) \sum_{i=0}^{b_1} p_{1i} \Psi_s(\vec{r}_{1i},t) \vec{c}_{1i} \Delta t \nonumber \\
&+& G^r_{s} \Psi_s(\vec{r};t) \sum_{i=1}^{b_2} p_{2i} \Psi_s(\vec{r}_{2i},t) \vec{c}_{2i} \Delta t \\ \label{eq:force}
\vec{F^X}_s(\vec{r}_i;t) &=& \frac{G_{AB}}{\rho_0} \rho_s(\vec{r};t) \sum_{i=0}^{b_1} w_i \rho_{s'}(\vec{r}_i;t) \vec{c}_i\Delta t \nonumber
\end{eqnarray}
In the above, the indices $k=1,2$ refer to the first and second Brillouin zones in the lattice
(belts, for simplicity), $\vec{c}_{ki}$, $p_{ki}, w_i$ are the corresponding discrete speeds and 
associated weights. $G_{AB} \equiv G_{ss'} = G_{s's}$, $s' \ne s$, 
is the cross-coupling between species, $\rho_0$ a reference density to be defined shortly and,
finally, $\vec{r}_{ki} \equiv \vec{r}+\vec{c}_{ki} \Delta t$ are the displacements
along the $i$-the direction in the $k$-th belt.
These interactions are sketched in Figure \ref{Fig:1}.
Note that positive(negative) $G$ code for repulsion(attraction) respectively.
Our model is reminiscent of the potentials
used to investigate arrested phase-separation and structural
arrest in charged-colloidal systems \cite{SCIORTINO}, and also bears similarities to
the NNN (next-to-nearest-neighbor) frustrated lattice spin models \cite{SETHNA}.
As compared with lattice spin models, in our case a high lattice connectivity is required 
to ensure compliance with macroscopic non-ideal hydrodynamics, particularly the isotropy of
potential energy interactions, which lies at the heart of the complex rheology to be discussed in this work.
To this purpose, the first belt is discretized with $9$ speeds ($b_1=8$), while the second 
with 16 ($b_2=16$) for a total of $b=24$ connections.
The weights are chosen in such a way as to fulfill the following normalization
constraints \cite{Chi_08}: 
$\sum_{i=0}^{b_1} w_i = \sum_{i=0}^{b_1} p_{i1} + \sum_{i=0}^{b_2} p_{i2} = 1$;
$\sum_{i=0}^{b_1} w_i c_i^2 = \sum_{i=0}^{b_1} p_{i1} c_{i1}^2 + \sum_{i=0}^{b_2} p_{i1} c_{i2}^2= c_s^2$,
$c_s^2=1/3$ being the lattice sound speed.
The pseudo-potential $\Psi_s(\vec{r})$ is taken in the form first suggested by Shan and Chen \cite{SC_93},
$\Psi_s[\rho] = \sqrt{\rho_0} (1-e^{-\rho/\rho_0})$, where $\rho_0$ marks the density value 
at which non ideal-effects come into play.
Following \cite{Sbr_07},  Taylor expansion of (\ref{FORCE}) to fourth-order in $\Delta t$ delivers the   
non-ideal pressure tensor $P_{\alpha \beta}(\vec r;t)$, namely
\begin{equation}
\label{PAB}
P_{\alpha \beta} = [c_s^2 \rho +  \frac{1}{2}c_s^2 G_{A1} \Psi_A^2 + G_{B1} \Psi_B^2+ c_s^4 \Pi] \delta_{\alpha \beta} - c_s^4\gamma_{\alpha \beta}
\end{equation}
where greek indices run over spatial dimensions and:
\begin{eqnarray}
\nonumber
&&\Pi = \Sigma_{s=A,B}G_{s2}[\frac{1}{4} (\nabla \Psi_s)^2 - \frac{1}{2} \Psi_s \Delta \Psi_s]+  \\
&& \frac{G_{AB}}{\rho_0} [ \rho_A \Delta \rho_B + \rho_B \Delta \rho_A - \nabla \rho_A \nabla \rho_B]  \\
\label{P1}
\nonumber
&&\gamma_{\alpha \beta} = \Sigma_{s=A,B} G_{2s} \partial_{\alpha} \Psi_s \partial_{\beta} \Psi_s + \\
&& \frac{G_{AB}}{2\rho_0} 
(\partial_{\alpha} \rho_A \cdot \partial_{\beta} \rho_B + \partial_{\alpha} \rho_B \cdot \partial_{\beta} \rho_A )
\label{P2}
\end{eqnarray}
In the above equations, we have introduced the 
effective couplings $G_{s1}=G_s^a+G_s^r$ and $G_{s2}=G_{s1}+ \frac{12}{7} G_{s2}$, $s=A,B$, respectively.
The non ideal  pressure splits into a local (bulk)  and 
non-local (surface) contributions, which fix the surface tension $\sigma$ of the model. 
It is crucial to appreciate that the value of $\sigma$ can be tuned by
changing the reference density $\rho_0$. 
The repulsive  intra-species  force $\vec{F}^r_s$ (proportional to $\sqrt{\rho_0}$)
 acts against the  inter-species repulsive force $\vec{F}^X$ (proportional to $1/\rho_0$). 
Thus, for small $\rho_0$, $\vec{F}^X$ dominates and a complete separation
 between the two fluids is expected. This is the case of large and positive $\sigma$. 
On the other hand, for large $\rho_0$, $\sigma$ becomes smaller and even negative. 
In figure \ref{Fig:2} (left panel), we show  $\sigma$ as a function
of the reference density $\rho_0$, as obtained through a standard Laplace test on
a single bubble configuration. As anticipated, the surface tension decreases at increasing
$\rho_0$ and becomes negative beyond a given threshold,  $\rho_0>0.71$. 
In this work we shall be concerned only with the
case of small and positive $\sigma$. 
In the following, we shall discuss numerical simulations with random initial conditions for the
two densities $\rho_A$ and $\rho_B$.
After a short transient,
the interfacial area reaches its maximum value  and progressively 
tends to decrease due to the effect of surface tension which drives the system 
towards a minimum-interface configuration. 
In the long term, this minimum-area tendency would lead to the complete separation 
between components A and B, with a single interface between two separate bulk components. 
However, such a tendency is frustrated (hence, strongly retarded)  by the
the complex  interplay between  repulsive (short-range inter-species and mid-range
intra-species) and attractive (short-range intra-species) interactions. The final result is a 
rich configurational dynamics of the density field, as the one shown in figure (\ref{Fig:2}) right panel.
\begin{figure}
\vspace{-0.5cm}
\includegraphics[height=6cm,width=8cm]{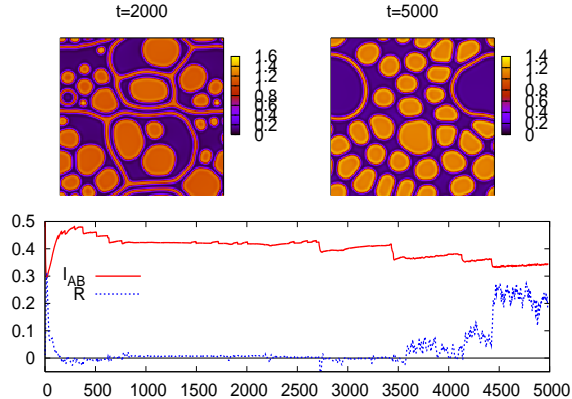}
\caption{The response function $R$ and the surface indicator $I_{AB}$ as a function of time, 
(expressed in units of $10^3$ LB time-steps). 
The forcing is $U_0=0.1$, the domain is $128^2$ and the other parameters are defined in the text. 
The sharp decrease of the response function in the initial stage indicates the structural arrest of the system,
associated with an anomalous enhancement of the flow viscosity, about four orders of magnitude above the
molecular value. In the left panel, cages are present, which manage to ``trap'' micro-structures inside. 
In the right panel, the cages break down and the system is now able to flow again.}
\label{Fig:3}
\end{figure}
In order to investigate the rheological properties of the composite LB fluid, we
put the system under a shear flow $U_x(x,y) = U_0 sin (ky)$, $U_y=0$, 
and measure the response function $R=\frac{\bar{U}}{U_0}=\frac{\nu_0}{\bar{\nu}},$
where $\bar{U}=\sum_yU(x,y)/N_y$ and $\bar{\nu}$ defines the effective viscosity. 
Under normal flow conditions, $R=1$, so that $R\ll1$ provides a
direct signal of enhanced viscosity and eventually, structural arrest.
The main coupling parameters are
$G^a_{A} = -12.55$
$G^r_{A} = +11.80$
$G^a_{B} = -11.70$
$G^r_{B} = +10.95$
$G_{AB} = +0.58$, and $\rho_0=0.7$. 
These parameters correspond to both species in the dense phase, with no phase transition,
hence they can be regarded as descriptive of
glassy micro/nanoemulsions, namely a dispersion of liquid within
another, immiscible liquid. 
By letting each species undergo phase transitions between a dense and light
phase, the same model could describe foamy materials as well.
The simulations are performed mostly  on a grid $128^2$ (except the one reported in figure (\ref{Fig:2})
up to $5\times 10^6$ LB time-steps.
\begin{figure}
\vspace{-2cm}
\includegraphics[scale=0.9]{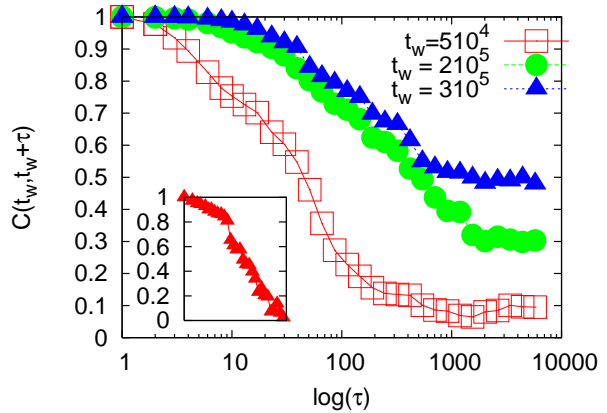}
\caption{
Ageing of the system.
Correlation function corresponding to different waiting times $t_w$ 
($t_w = 5~10^4 $ , red squares,  $t_w = 2~10^5$,  green circles and $t_w=310^5$, blue triangles)
 with shear stress $U_0 = 0.02$.
In the inset, we show the correlation function for $t_w=3~10^5$ and $U_0=0.03$:
with increasing shear stress the structural arrest disappears.
}
\label{Fig:4}
\end{figure}
In figure \ref{Fig:3}, we show the time evolution of the response $R$, as well as
an indicator of the interface area, $I_{AB} = \sum_{x,y} \nabla \rho_A \cdot \nabla \rho_B$.
From this figure, we appreciate a very dramatic drop of the flow speed in the initial
stage of the evolution, corresponding to a very substantial enhancement of the fluid
viscosity (about four orders). The system remains in this 'arrested' state for a very long
time,  over three millions timesteps, until it suddenly starts to regain its initial velocity 
through a bumpy dynamics, characterized by a series of sudden jumps~\cite{BARRAT}.
These viscosity jumps signal 'plastic events', whereby the system manages to
break the density locks (cages) which blocked the flow in the initial phase.
As a result, the system progressively regains its capability to flow.
These plastic events are also recorded by the time trace of the interface area $I_{AB}$, which exhibits
an alternate sequence of plateaux followed by sudden down-jumps, the latter being responsible for the
overall reduction of the interface area as time unfolds.
Visual inspection of the fluid morphology confirms this picture. 
In the top panels of figure \ref{Fig:3}, we show the density field in an arrested state at time
$t=2 \;10^5$ (top-left) and in a flowing state, $t=5 \;10^5$ (top-right).
The left figure clearly reveals the existence of "cages" in the density field configuration, which
entrap the fluid inside and consequently block its net macroscopic motion. 
Inter-domain relaxation can only take place in response to 'global moves' of the density field, i.e. the "cage" rupture. 

Due to the mesoscopic nature of the present model, the rupture of a single cage in the
LB simulation corresponds to a large collection of atomistic ruptures, and consequently
it leads to observable effects in terms of structural arrest the system.
To the best of our knowledge, this the first time that such an effect is 
observed by means of a mesoscopic lattice Boltzmann model. 

To be noted that the use of high-order lattices (24-speeds) is instrumental to this
program, since, by securing the isotropy of lattice tensors up to 8th order, such
lattice permits to minimize spurious effects on the non-ideal hydrodynamic 
forces acting upon the discrete lattice fluid~\cite{Sbr_07}. 
We next inspect another typical phenomenon of soft-glassy matter, namely ageing.
To this purpose, following upon the spin-glass literature \cite{CAV}, we define 
the order parameter $\phi \equiv (\rho_A-\rho_B)$ and compute its {\it overlap}, defined through the 
autocorrelation of this order parameter:
\be
\label{CORRE}
C(t_w,\tau) = \frac{\langle \sum_{x,y} \phi(x,y;t_w) \phi(x,y;t_w+\tau) \rangle}
{\langle \sum_{x,y} \phi(x,y;t_w) \phi(x,y;t_w) \rangle}
\ee
where $t_w$ is the waiting time, $\tau$ is the time lapse between the two
density configurations and brackets stand for averaging over an ensemble of realizations.
In figure \ref{Fig:4}, we show the correlation function corresponding to different waiting times $t_w$ 
($t_w = 5 \;10^4 $ , red squares,  $t_w = 2 \; 10^5$,  green circles and $t_w=310^5$, blue triangles)
 for shear stress $U_0 = 0.02$.
From this figure, ageing effects are clearly visible, in the
form of a slower than exponential decay of the correlation function, which saturates
to a non-zero value in the long-time limit (broken ergodicity).
In the inset of the same figure, we show the correlation function for $t_w=3 \; 10^5$ and $U_0=0.03$:
with increasing shear stress the structural arrest disappears,
which is one of the most distinctive features of flowing soft-glassy materials~\cite{Cou_02}.
The main advantage of the present lattice mesoscopic approach is to give
access to hydrodynamic scales within a very affordable computational budget.
With reference to micro-emulsions (say water and oil), we 
note that the presence of surfactants usually gives rise to microscopic structures
of the order of $50$ nm in size~\cite{Wu_02}. These can be likened to the 'blobs'
observed in our simulations. 
With reference to liquid water, we have $\nu \sim 10^{-6}$ ($m^2/s$), which can be used to obtain a 
physical measure of the time step $\Delta t \sim 4$ ps, about three orders of magnitude larger than the typical
timestep used in Molecular Dynamics.
As a result, a five-million time-step LB simulation spans about 
$20$ microseconds in physical time.
Since the present LB method is easily amenable to parallel computing, parallel
implementations will permit to track the time evolution
of three-dimensional micro-emulsions of tens of microns in size, over time 
spans close to the millisecond, i.e. at space-time scales of hydrodynamic relevance.
Summarizing, we have provided the first numerical evidence that a two-species mesoscopic
lattice Boltzmann model with mid-range repulsion between like-molecules
and short-range repulsion between different ones, is capable of 
reproducing many distinctive features of soft material behaviour, such as 
slow-relaxation, anomalous enhanced viscosity, caging effects and aging under shear.
The present lattice kinetic model caters for this very rich physical picture  
at a computational cost only marginally exceeding the one for a simple fluid. 
As a result, it is hoped that it can be used as an alternative/complement
to MonteCarlo and/or Molecular Dynamics, for future investigations
of the {\it non-equilibrium} rheology of a broad class of flowing disordered materials, such as
microemulsions, foams and slurries, on space and time scales of experimental interest.

SS wishes to acknowledge financial support from the project INFLUS (NMP3-CT-2006-031980).
SC wishes to acknowledge financial support from the ERG EU grant and COMETA.
Fruitful discussions with L. Biferale, D. Nelson, G. Parisi and F. Toschi are kindly acknowledged.
The authors are thankful to A. Cavagna for critical reading of this manuscript.

\end{document}